\documentclass[3p,times,twocolumn]{elsarticle}

\usepackage{ecrc}

\volume{00}

\firstpage{1}

\journalname{Nuclear Physics B Proceedings Supplement}

\runauth{L. Amati et al.}

\jid{nuphbp}

\jnltitlelogo{Nuclear Physics B Proceedings Supplement}

\usepackage{amssymb}

\usepackage[figuresright]{rotating}

\begin{document}

\begin{frontmatter}


\title{The LOFT contribution to GRB science}
 \cortext[cor1]{Email: amati@iasfbo.inaf.it}
 \fntext[list]{http://www.iasfbo.inaf.it/~amati/loftgrb}

\dochead{}

\title{The LOFT contribution to GRB science}


\author[iasfbo]{L. Amati\corref{cor1}}
\author[iaps]{E. Del Monte}
\author[asdc]{V. D'Elia}
\author[asdc]{B. Gendre}
\author[iasfmi]{R. Salvaterra}
\author[mporzio]{G. Stratta}
\author[]{on behalf of the LOFT/GRB team\fnref{list}}

\address[iasfbo]{INAF - IASF Bologna, via P. Gobetti 101, 40129 Bologna, Italy}
\address[iaps]{INAF - IAPS, via Fosso del Cavaliere 100, 00133 Roma, Italy}
\address[asdc]{ASI Science Data Center (ASDC), via Galileo Galilei, 
00044 Frascati, 
 Italy}
\address[iasfmi]{INAF - IASF Milano, via E. Bassini 15, I-20133 Milano, Italy}
\address[mporzio]{INAF - Osservatorio Astronomico di Roma, via Frascati 33, 00040 Monte Porzio Catone (RM), Italy}

\begin{abstract}

LOFT is a satellite mission currently in Assessment Phase for the
ESA M3 selection. The payload is composed of the Large Area
Detector (LAD), with 2--50 keV energy band, a peak effective area
of about 10 m$^2$ and an energy resolution better than 260 eV, and
the Wide Field Monitor (WFM), a coded mask imager with a FOV of several
steradians, an energy
resolution of about 300 eV and a point source location accuracy of
1 arcmin in the 2--30 keV energy range.
Based on preliminary considerations and estimates, we show how
the scientific performances of the WFM are particularly suited to
investigate some of the most relevant open issues in the study of GRBs: the
physics of the prompt emission, the spectral absorption features by
circum-burst material (and hence the nature of the progenitors),
the population and properties of XRFs, and the detection and rate of high--z GRBs.
Measurements of the early afterglow emission with the Large Area Detector (LAD)
may also be possible depending on the mission slewing capabilities and TOO
observations policy.

\end{abstract}


\end{frontmatter}


\begin{table*}[t]
\caption{Main characteristics of the LOFT Wide Field Monitor (WFM) \cite{Brandt12,Feroci12b}.
}
\centerline{\includegraphics[width=1.4\columnwidth]{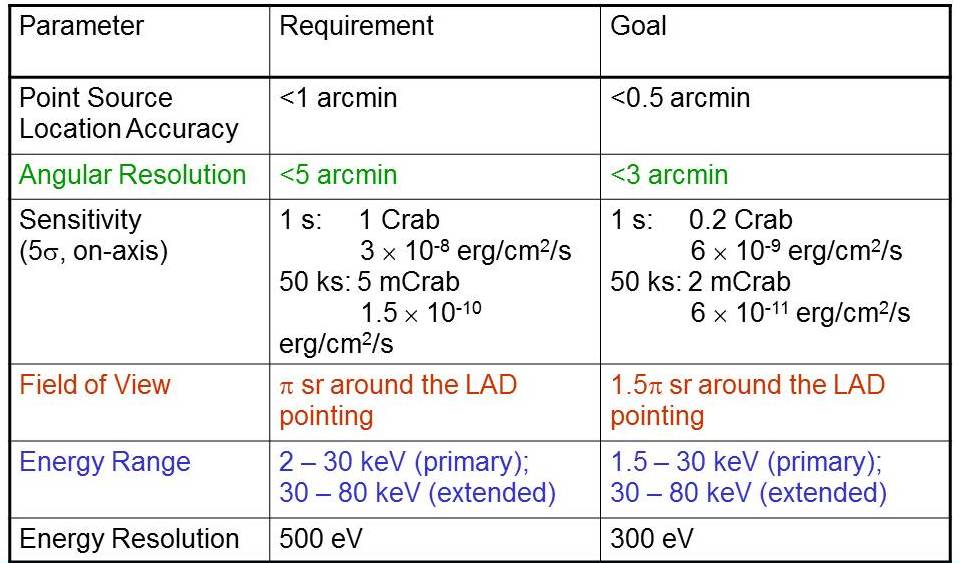}}
\label{ta1}
\end{table*}

\section{Introduction}

Despite of huge advances occurred in the last 10 years, the Gamma--ray
Bursts (GRB) phenomenon is still far to be fully understood. Open issues
include: physics and geometry of the prompt emission, unexpected
early afterglow phenomenology (e.g. plateau phase, flares), identification
and understanding of sub-classes of GRBs (short/long, X--Ray Flashes, 
sub--energetic
GRBs), GRB/SN connection, nature of the inner
engine, cosmological use of GRBs, and more. See, e.g., \cite{Meszaros06,Gehrels09,Zhang11} 
for recent reviews.

The Large Observatory For X--ray Timing, LOFT \cite{Feroci12a,Feroci12b}, was selected in 2011 by the
European Space Agency as one of the four Cosmic Vision M3 candidate
missions to compete for a launch opportunity at the start of the 2020s.
The current schedule of the ESA/M3 programme foresees the end of the assessment 
study by the end of 2013 and the final down--selection in the first months of 2014.

Thanks to an innovative design and the development of large-area
monolithic silicon drift detectors, the Large Area Detector (LAD) on board
LOFT will operate in the 2--30 keV range (up to 80 keV in expanded mode) with
a FOV collimated to $\sim$1$^{\circ}$,
and achieve an effective area of $\sim$10 m$^2$ at 8 keV, a time resolution of $\sim$10
$\mu$s and a spectral resolution of $\sim$200--260 eV (FWHM at 6 keV)\cite{Bozzo11,Zane12}. 
These capabilities will allow LOFT 
to investigate with unprecedented sensitivity the rapid X--ray flux and spectral variability that directly probes 
the motion of matter
down to distances very close to black holes and neutron stars, as well as the physical state of 
ultra-dense matter \cite{Feroci12a}. Such measurements are
efficient diagnostics of the behavior of matter in the presence of strong gravitational fields, 
where the effects predicted
by General Relativity are more relevant, and the physics of matter at densities in excess of that 
in atomic nuclei, determining its
equation of state and composition. 

In addition to the LAD, the LOFT payload will include a Wide Field Monitor
(WFM)\cite{Brandt12}, which will monitor more
than half of the LAD-accessible sky (approximately 1/3 of the whole sky) simultaneously at any time,
will operate in the same energy range as the LAD with an energy resolution of
$\sim$300 eV, providing information about source status 
(flux variability and energy
spectrum), as well as arc--minute positioning (Tab.~1). 
With such wide angle sky monitoring, the WFM will also provide long term
histories of the target sources, thus facilitating both the LAD observations and a series of wider 
science goals.

Based on preliminary considerations and estimates, we show that LOFT, possibly in combination with 
other GRB experiments flying at the same epoch, 
will give us useful and
unique clues to some of these still open issues in this field, through:
a) measurements of the prompt emission down to $\sim$2 keV and $\sim$arcmin localization with the Wide
Field Monitor (WFM);
b) measurements of the early afterglow emission with the Large Area Detector (LAD).

As detailed below, the partly unprecedented characterization of the GRB X--ray prompt emission, 
joined with
source location accurate enough for optical follow-up, by the WFM will allow to investigate the 
properties of the
circum-burst environment, thus getting further clues on the nature of the progenitors, to provide a more
stringent tests for the emission mechanisms at play, to increase the detection rate of high--z GRBs w/r to
previous missions, to shade light on the population of X--ray Flashes (XRFs) and sub-energetic events. While
the GRB science with the WFM will come 'for free', provided the optimization of trigger logic, 
data modes
and prompt (within a few tens of s) transmission of trigger time and position to ground (which is
presently in the baseline of the mission \cite{Feroci12b,Brandt12}), the contribution from the LAD, consisting in the possible further characterization of
the 'plateau phase' of the early X-ray afterglow emission, will critically depend on the follow-up capabilities and
policy of the mission.

\section{GRB science with the LOFT/WFM}

The WFM will achieve scientific goals of fundamental importance and not fulfilled by GRB experiments
presently flying (e.g., Swift\cite{Gehrels04}, Konus/WIND\cite{Aptekar95}, 
Fermi/GBM\cite{Meegan09}) and future approved missions (SVOM\cite{Godet12}, 
UFFO\cite{Grossan12}).
These can be summarized as follows:
\begin{itemize}
\item measurement of the GRB spectral shape and its evolution down to about 2 keV in photon energy which is
crucial for testing models of GRB prompt emission (still to be settled despite the considerable amount of
observations; see, e.g., \cite{Zhang02});

\item detection and study of transient X--ray absorption column / features for a few 
bright GRBs per
year (Fig.~1). These measurements are of paramount importance for the understanding of the properties of the
Circum--Burst Matter (CBM) and hence the nature of GRB progenitors (still a fundamental open issue in the
field). In addition, as demonstrated by BeppoSAX \cite{Amati00}, the detection of transient X--ray spectral features can
allow the determination of the GRB redshift to be compared, when it is the case, with that determined from the
optical/NIR lines;

\item to provide a substantial increase (with respect to the past and current missions) in the detection rate of X--Ray
Flashes (XRF), a sub--class of soft / ultra-soft events which could constitute the bulk of the GRB
population (e.g., \cite{Pelangeon08}) and still have to be explored satisfactorily;

\item to extend the GRB detection up to very high redshift (z $>$ 8) GRBs (Fig.~2), 
which is of fundamental importance for
the study of evolutionary effects, the tracing of star formation rate, ISM evolution, and possible unveiling of
population III stars (e.g., \cite{Salvaterra10,Woosley06});

\item to provide fast ($<$30s) and accurate ($\sim$1 arcmin) location of the detected GRBs to allow their prompt multi--wavelength follow-up
with ground and space telescopes, thus leading to the identification of the optical counterparts and/or host
galaxies and to estimate the redshift, a fundamental measurement for the scientific goals listed above,
comparing it with that determined from X-ray absorption lines.

\end{itemize}

\begin{figure}[t]
\includegraphics[width=1.1\columnwidth]{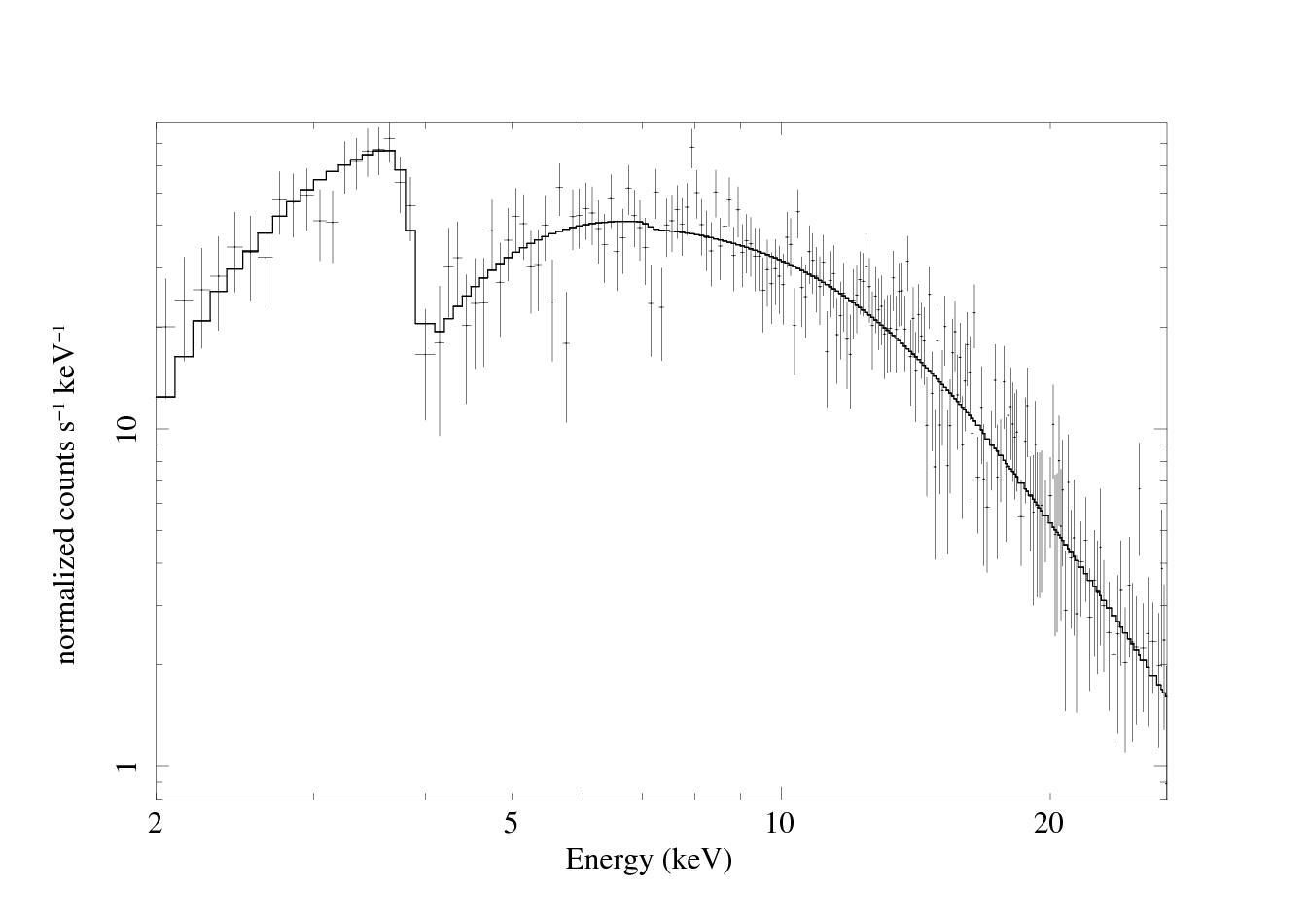}
\caption{
The transient absorption edge detected in the first 13 s of GRB 990705 \cite{Amati00} as
would be measured by the LOFT/WFM.
}
\end{figure}

\begin{figure}[t]
\includegraphics[width=\columnwidth]{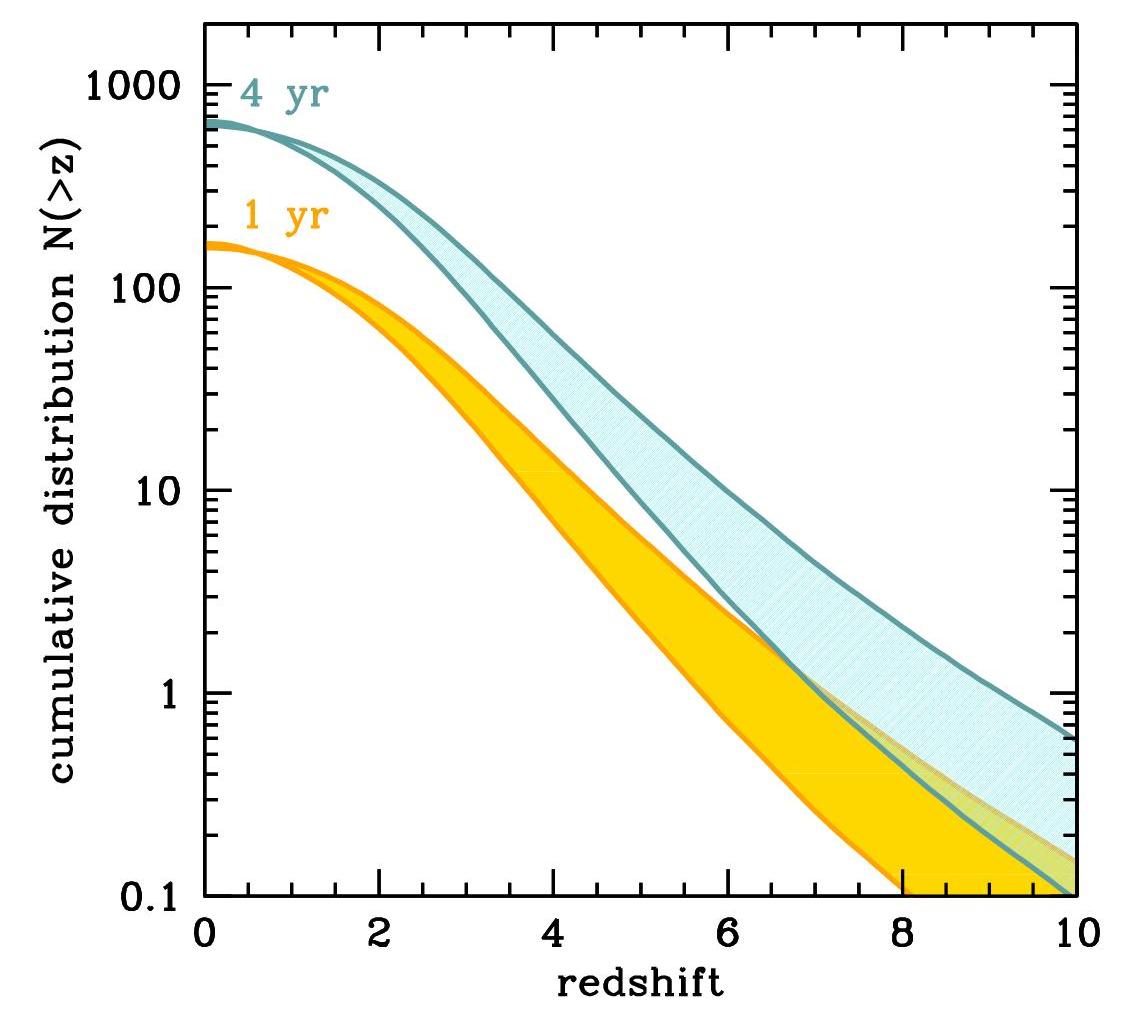}
\caption{
Expected GRB+XRF detection rate by the LOFT/WFM as a function of
redshift . The width of the strips reflect the uncertainties in the GRB luminosity function and its evolution with redshift
(see \cite{Salvaterra10}).
}
\end{figure}

\section{Possible further contribution by the LOFT/LAD}

The GRB science that could be done with the LAD is less 'automatic' and is strongly dependent on the 
time
that will be required to start a TOO observation. Possible GRB science that could be performed by the LAD
include:
\begin{itemize}

\item investigating the plateau phase of the early X--ray afterglow and its transition to the 'normal decay' (Fig.~3), 
which is of
high importance e.g. for the search of signatures of energy injection by a magnetized NS and to better
investigate the correlation between time and luminosity of the transition from plateau to normal decay (e.g.,
\cite{Zhang07});
\item searching for emission lines, expected by theoretical models in case of highly metal enriched circum--burst
environment and detected in a few cases by BeppoSAX, Chandra and XMM (e.g., \cite{Piro00}) but
not by Swift/XRT;
\item complementing the observations of the prompt emission by the WFM by exploiting the transparency of the
collimator at energies $>$ 30--40 keV.

\end{itemize}

\begin{figure*}[t]
\includegraphics[width=\columnwidth]{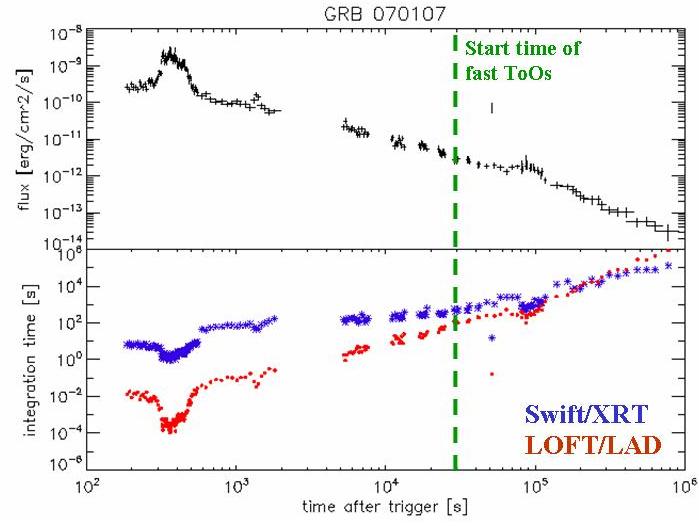}
\includegraphics[width=\columnwidth]{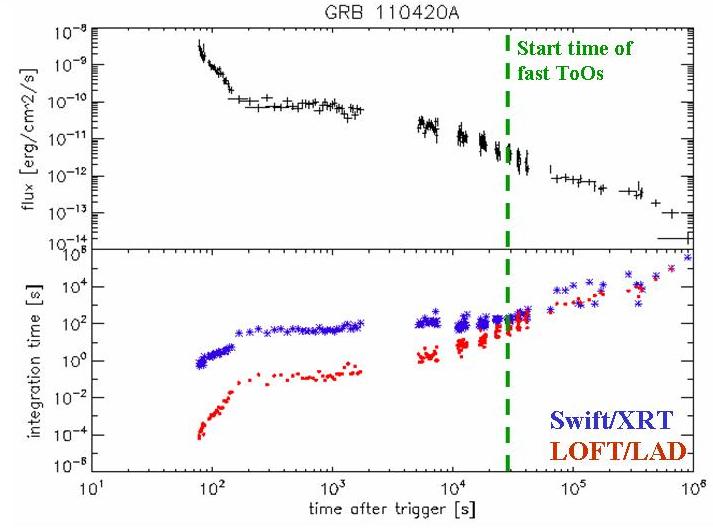}
\caption{
Top panels: light curves of the early afterglow of two GRBs observed by Swift. Bottom panels: time
required to reach a 5$\sigma$ significance in a given time bin for the Swift/XRT and the LOFT/LAD. As can be seen,
the LOFT/LAD would be more sensitive than the Swift/XRT up to about 8 hours from the GRB onset.
}
\end{figure*}

\section{LOFT GRB science in the $>$2020 context}

No past, present or future (e.g., SVOM, UFFO) GRB experiment has such a combination of low energy threshold and
high energy resolution (and wide FOV), which will make the LOFT/WFM unique also for GRB science. For
instance, the BeppoSAX/WFC or the HETE--2/WXM had a low energy threshold around 2 keV, but with much
worse energy resolution and smaller effective area (and smaller FOV). SVOM and UFFO ($>$2018) will have a
low energy threshold of $\sim$5 keV and with significantly worse effective area and energy resolution w/r to
LOFT/WFM at energies $<$10 keV.
In addition, in particular thanks to the capability of promptly transmitting the GRB trigger
and a
position with a few arcmin uncertainty \cite{Brandt12},
LOFT will:
a) continue the fundamental 'service' to the astrophysical community, carried out presently by Swift (and
possibly in the 2018-2022 time frame by SVOM and/or UFFO) of providing detection, prompt localization and temporal / spectral
characterization of GRBs, thus allowing their follow--up and multi--wavelength studies with the best telescopes
operating in the $>$2020 time line (e.g., LSST, SKA, CTA, eROSITA, maybe XMM, Chandra, etc.);
b) complement simultaneous observations by GRB experiments flying on other satellites by providing low
energy extension and GRB position, in a way similar to what is presently done, e.g., by joining data from Swift,
Fermi and Konus/WIND.

In addition, the study of GRBs, because of their complex phenomenology and extreme energetics, 
the
association of at least a fraction of them with core--collapse SNe, their redshift distribution extending up to at
least z$\sim$8--9, is relevant to different fields of astrophysics, ranging from plasma and black--hole physics to
cosmology. In particular, the LOFT measurements summarized above will provide not only a step forward in
the understanding of GRB physics, progenitors and sub--classes, but, also in combination with observations by
the best instrumentation at all wavelengths, will shade light on fundamental topics like the evolution of the star
forming rate and of the galactic ISM up to the epoch of re-ionization, the first generation (pop III) stars, the
understanding of the diversity and rate of core-collapse SNe, the measurement of cosmological parameters.

Finally, the GRB science already involves, and will continue to involve at the time of LOFT, observers and
instrumentation from different communities: optical/IR robotic telescopes (prompt detections and localization of
the optical counterparts), major optical/IR telescopes like VLT, Gemini, Hubble (identification, redshift and ISM
of the host galaxies, optical afterglow decay and jet signatures), radio telescopes like VLA (afterglow modeling
and energetics) observations in the TeV range by Cherenkov telescopes (challenging emission models).





\end{document}